\documentclass[a4paper,fleqn,usenatbib]{mnras}

\usepackage{newtxtext,newtxmath}
\usepackage[T1]{fontenc}
\usepackage{ae,aecompl}

\usepackage{graphicx}	% Including figure files
\usepackage{amsmath}	% Advanced maths commands
\usepackage{cleveref}
\usepackage{pgf}

\usepackage{subfig}
\usepackage[utf8]{inputenc}		% accented letters
\usepackage{pifont}
\usepackage{romannum}
\usepackage[binary-units,range-units=single]{siunitx}
\DeclareSIUnit\parsec{pc}
\newcommand{\Msol}{{\textrm{M}}_\odot}

\title{Detection of a dark matter subhalo in the strongly lensed system PJ011646}

\author[A. Amvrosiadis et al.]{A. Amvrosiadis$^{1, 2}$\thanks{E-mail: aristeidis.amvrosiadis@physics.ox.ac.uk},
J. W. Nightingale$^{3}$,
Q. He$^{4}$, 
A. Robertson$^{5}$,
S. Cole$^{2}$, 
C. S. Frenk$^{2}$, 
S. Lange$^{5}$, \newauthor
R. Massey$^{2, 7}$, 
M. von Wietersheim-Kramsta$^{2, 7}$,
X. Cao$^{8}$, 
R. Li$^{9}$,
S. Li$^{10,11,9}$,
K. Wang$^{10,11,9}$,
X. Ma$^{10,11,9}$, \newauthor
L. W. H. Fung$^{2}$
\vspace{4mm}\\
$^{1}$ Sub-department of Astrophysics, Department of Physics, University of Oxford, Denys Wilkinson Building, Keble Road, Oxford, OX1 3RH, UK \\
$^{2}$ Institute for Computational Cosmology, Department of Physics, Durham University, South Road, Durham DH1 3LE, UK \\
$^{3}$ School of Mathematics, Statistics and Physics, Newcastle University, Newcastle upon Tyne, NE1 7RU, UK \\
$^{4}$ Kapteyn Astronomical Institute, University of Groningen, PO Box 800, NL9700 AV Groningen, The Netherlands  \\
$^{5}$ Carnegie Observatories, 813 Santa Barbara Street, Pasadena, CA 91101, USA \\
$^{6}$ Department of Physics, Chinese University of Hong Kong, Shatin, N.T., Hong Kong \\
$^{7}$ Centre for Extragalactic Astronomy, Department of Physics, Durham University, South Road, Durham DH1 3LE, UK \\
$^{8}$Institute for Astrophysics, School of Physics, Zhengzhou University, Zhengzhou, 450001, China\\
$^{9}$ School of Physics and Astronomy, Beijing Normal University, Beijing 100875, China \\
$^{10}$ School of Astronomy and Space Science, University of Chinese Academy of Sciences, Beijing 100049, China \\
$^{11}$ National Astronomical Observatories, Chinese Academy of Sciences, 20A Datun Road, Chaoyang District, Beijing 100101, China \\
}

\pubyear{???}

\begin{document}
\pagenumbering{arabic}
\label{firstpage}
\maketitle

% Abstract
\begin{abstract}

\noindent We present a strong lensing analysis of the system PJ011646 using high-resolution ($\sim$0.1 arcsec) Atacama Large Millimeter/submillimeter Array (ALMA) dust-continuum observations to test for the presence of dark matter substructures. The lens mass distribution is modelled with an elliptical power law and third- and fourth-order multipoles (PL+MP; $m=3,4$), plus external shear. The multipoles have amplitudes of $\simeq$1.5 per cent of the convergence, consistent with nearby early-type galaxies, and improve the fit by $\Delta\ln Z = 52.1$ relative to a pure PL model. Using this best-fitting macromodel, we perform a grid-based subhalo search in the image plane, parametrising the perturber as a spherical NFW. A subhalo in two locations in the image plane improves the fit by $\Delta\ln Z>10$. Both correspond to the same location in the source plane, so they are partially degenerate; follow-up analysis suggests that only one is physically real. This is a subhalo of mass $ M_{200} ={2.78}_{-0.66}^{+0.43} \times 10^{10} \, \Msol$ and concentration $c_{200} = 30_{-7}^{+5}$, detected at $\sim$5.8$\sigma$ significance (relative to the PL+MP). This concentration is consistent with that expected for a typical tidally stripped Navarro-Frenk-White subhalo. The enclosed projected mass is most tightly constrained within a radius of 2 kpc, where we infer $M_{\rm sub} = {3.57}_{-0.14}^{+0.16}\times 10^9 \, \Msol$. From grid cells consistent with no detection ($\Delta \ln Z < 10$), we derive limits on the minimum subhalo mass that could have been detected at $3\sigma$ significance, finding $M_{200} \approx 8 \times 10^{8} \, \Msol$ in the most sensitive regions of the lensed arcs. This demonstrates that ALMA continuum imaging at sub-arcsecond resolution can probe dark matter substructure in a mass regime where cold and warm dark matter models predict different abundances of subhalos.

\end{abstract}


%%%%%%%%%%%%%%%%%%%%%%%%%%%%%%%%%%%%%%%%%%%%%%%%%%%%%%%%%%%%%%%%%%%%%%%%%%%
% SECTION
%%%%%%%%%%%%%%%%%%%%%%%%%%%%%%%%%%%%%%%%%%%%%%%%%%%%%%%%%%%%%%%%%%%%%%%%%%%
\section{Introduction} \label{sec:section_1}

The nature of dark matter, or more specifically the identity of the dark matter particle, is a fundamental unsolved problem in cosmology. The widely accepted cold dark matter (CDM) model assumes that the dark matter particle was either born out of thermal equilibrium (an axion) or had very small thermal motions at early times. This model has been remarkably successful at accounting for a number of observations of the cosmic large-scale structure, from the spatial inhomogeneities seen in the temperature of the cosmic microwave background radiation \citep[e.g.][]{2020A&A...641A...6P} to the clustering of galaxies on large scales \citep[e.g.][]{2017MNRAS.470.2617A}. At the core of this theory is the formation of self-bound dark matter halos by a hierarchical process driven by gravitational instability \citep{Frenk_1988, 2008MNRAS.391.1685S, 2020Natur.585...39W}; {for a review see} \citealt{FW_2012}.

A fundamental prediction of CDM is that the halo mass function (HMF), defined as the number density of halos per logarithmic mass interval, follows an approximate power-law form at low masses, $\mathrm{d}n/\mathrm{d}\log M \propto M^{-\alpha}$, with $\alpha \sim 1$, and exhibits an exponential cut-off at the high-mass end \citep[e.g.][]{2001MNRAS.321..372J}. The cold nature of the dark matter means that halos can form over a huge range of masses, from that of the Earth to that of a rich galaxy cluster. By contrast, thermal motions of particles at early times in warm dark matter (WDM) models induce a cutoff in the HMF at a scale that depends inversely on the particle mass \citep[e.g.][]{2012MNRAS.420.2318L}. These two possibilities, CDM and WDM, can therefore be distinguished by the abundance of low-mass halos. WDM predicts a suppression in the number of low-mass halos below a characteristic cut-off scale, but {\em subhalos} can be significantly stripped by gravitational tides after infall and reach masses well below this scale\footnote{We will generally use the term ``subhalo''  although in most cases we do not know if the perturber is a subhalo in the lens or a field halo along the line-of-sight}. Consequently, the detection of a single low-mass subhalo does not unambiguously rule out WDM; instead, meaningful constraints require statistical measurements of the subhalo population and comparison with theoretical expectations.

Lower limits on the mass of a WDM particle have been derived from the measured abundance of dwarf galaxies, including satellites of the Milky Way \citep[][and references therein]{Newton_2021} and the Lyman-$\alpha$ forest \citep{Viel_2005}. These studies rule out most candidates for which the cut-off mass in the HMF exceeds the critical mass above which gas can cool by atomic processes and form stars \citep[e.g.][]{2020MNRAS.498.4887B}\footnote{This is not strictly true for particles such as sterile neutrinos for which the HMF cut-off depends not only on the particle mass but also on additional parameters \citep {Newton_2025}.}. Dark halos below that mass, however, can be detected through their strong gravitational lensing effects.  A strong lens forms when light from a background galaxy is deflected by a foreground galaxy aligned along the line-of-sight to an observer and forms multiple images stretched in arc-like configurations, such as an Einstein ring.  The strength of light bending is directly proportional to the total mass interior to the multiple images. Low-mass halos in the vicinity of the lens halo, or along the line-of-sight to it, induce localized small-scale perturbations to the lensed images. By modelling the lensed emission it is possible to infer, or exclude, the presence of these dark matter subhalos. The best limits to date on the mass of a WDM particle come from combining strong lensing constraints with those from satellites and the Lyman-$\alpha$ forest \citep{Enzi_2021}. 

The first strong lensing detection of a subhalo was made in the galaxy SLACS J0946+1006, using Hubble Space Telescope (HST) data \citep[][]{2010MNRAS.408.1969V}. Several independent studies have confirmed this detection using different modelling approaches \citep[e.g.][]{2021MNRAS.507.1662M, 2024MNRAS.528.7564B, 2024MNRAS.52710480N, 2025MNRAS.540..247E, 2025A&A...699A.222D, 2025MNRAS.543..540T, 2025arXiv250419177C, 2025ApJ...991L..53H}. The estimated mass of this subhalo is $\rm M_{200} \sim 10^{10} \, \Msol$ (assuming an NFW profile with a free concentration). While some studies found it to be unusually concentrated, in disagreement with CDM predictions, more recent work explicitly accounting for the faint light expected from such a massive subhalo shows that less concentrated solutions, that are consistent with CDM predictions, are preferred \citep[][]{2025ApJ...991L..53H}. 

The second subhalo detection was reported in JVAS B1938+666 from  adaptive optics data in the Keck telescope \citep[][]{2012Natur.481..341V}. The estimated mass is $\rm M_{200} \sim 10^{9} \, \Msol$. Recent re-analyses have reclassified this object as a foreground field halo along the line-of-sight rather than a subhalo of the main lens \citep[][]{2025MNRAS.543..540T}. This detection has been reproduced by \citet{2022MNRAS.515.4391S} and \citet{2025ApJ...991L..27L}, although the inferred properties (mass, concentration and redshift) differ significantly. 

A claimed detection in SDP.81 using ALMA data \citep[][]{2016ApJ...823...37H}, the first and, so far, the only one reported with this instrument, has since been shown to be a false positive caused by an inaccurate macro-model of the lensing galaxy \citep[][]{2025arXiv250802776S}.

\citet{2025NatAs.tmp..205P} recently reported a subhalo detection using very long baseline interferometry (VLBI), with an inferred enclosed projected mass within 80 pc of $\rm M_{\rm 2D} \sim 10^6 \, \Msol$, the lowest mass structure ever found through strong gravitational lensing. While this result is remarkable, only two strongly lensed radio jets have so far been discovered and observed with VLBI, underscoring the need for larger samples to establish the existence and abundance of such low-mass perturbers. A tentative subhalo detection has been reported in SPT2147–50 using JWST data \citep[][]{2025MNRAS.539..704L}, with an estimated mass of $\rm M_{200} \sim 7 \times 10^{10} \, \Msol$.

Alongside individual detections, non-detections are also valuable \citep[e.g.][]{2018MNRAS.481.3661V, 2019MNRAS.485.2179R, 2024MNRAS.52710480N}, as they allow upper limits to be translated into constraints on the WDM particle mass \citep[e.g.][]{2018MNRAS.481.3661V, Enzi_2021}. There are complementary probes to strong lensing distortions. Flux-ratio anomalies in quadruply lensed quasars currently provide the strongest astrophysical limits, implying a WDM particle mass of $m_{\rm WDM} > 8.4$ keV \citep[][]{2025arXiv251107765K, 2025arXiv251107513G}. This is significantly tighter than constraints derived from Milky Way satellite counts, which yield $m_{\rm WDM} > 2.0$ keV \citep{Newton_2021}. These approaches rely on different assumptions and systematics, highlighting the importance of multiple, independent methods for constraining the properties of the dark matter particles.

In the past few years, considerable effort has been devoted to improving the modelling techniques used in strong-lensing analyses \citep{2024MNRAS.532.2441H, 2025MNRAS.540..247E, 2025ApJ...981....2M}, assessing the limitations of existing datasets \citep{2022MNRAS.510.2480D, 2023MNRAS.521.2342O}, and characterizing the main sources of systematic uncertainty in subhalo searches \citep{2023MNRAS.518..220H, 2024MNRAS.528.1757O, 2024MNRAS.531.3431C}. The reliability of subhalo inferences hinges on how closely the adopted macro-model reflects the real  structure of the lens galaxy.  Early type galaxies (ETGs) are more complex than the simplified models typically used, and any mismatch can bias or mimic subhalo signals. 

Most galaxy-scale strong lensing studies assume an elliptical power-law (PL)  distribution for the total (dark + baryonic) mass in ETGs \citep{2010ARA&A..48...87T}. However, ETGs often exhibit features in their stellar light distributions that deviate from elliptical symmetry. As observational data have improved in resolution and sensitivity, such deviations have become increasingly apparent, exposing the limitations of simple elliptical models. For instance, PL models can fail to accurately focus light rays onto the source plane, leaving coherent wave-like patterns in the reconstructed images \citep{2022MNRAS.516.1808P, 2024CmPhy...7..286B}.

Several recent studies have emphasized that unaccounted for angular complexity in the lens potential can masquerade as the signal of a dark matter subhalo. In this situation, the inclusion of a subhalo is statistically preferred not because a true perturbing mass exists, but because the assumed subhalo acts to “correct” imperfections in the macro-model \citep[e.g.][]{2023MNRAS.518..220H, 2024MNRAS.52710480N}. This degeneracy between subhalo perturbations and higher-order structure in the smooth mass distribution represents one of the dominant sources of systematic uncertainty in subhalo detection. The issue is not confined to substructure studies alone—it also affects time-delay cosmography and quasar flux-ratio analyses, where accurate modelling of the lens potential is equally critical \citep[e.g.][]{2022A&A...663A.179V, 2024MNRAS.531.3431C}.

\begin{figure*}
\centering
\includegraphics[width=0.975\textwidth]{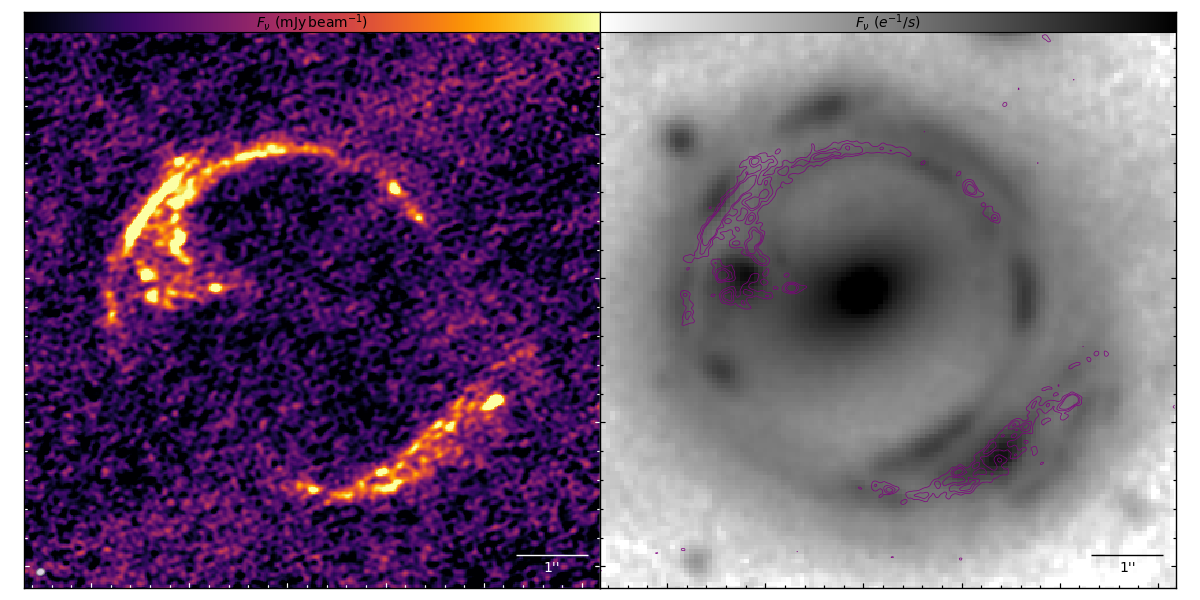} 
\caption{\textit{(Left panel):} ALMA dust continuum image Band 7, for PJ011646 using the natural weighting scheme. The synthesized beam is shown in the bottom left corner as a white ellipse. This image is only intended for visualization since the modelling is carried out directly in the uv-plane. \textit{(Right panel):} HST/WFC3 stellar continuum image in the F160W filter. Purple contours show the dust continuum emission at 3, 5 and 7$\sigma$.}
\label{fig:data}
\end{figure*}

To mitigate these shortcomings, several approaches have been developed to introduce additional flexibility into the macro-model. One approach is to include higher-order Fourier terms (“multipoles”) in the mass model of the lensing galaxy \citep[e.g.][]{2016ApJ...823...37H, 2022MNRAS.516.1808P, 2022A&A...659A.127V, 2024MNRAS.528.7564B, 2024A&A...688A.110S, 2024MNRAS.531.3431C}. This captures small but coherent deviations from elliptical symmetry, consistent with observations of local ETGs that exhibit similar multipole structures in their light distributions \citep[e.g.][]{2018ApJ...856...11G, 2025MNRAS.540.3281A}. An alternative approach is to use composite models that explicitly separate baryonic and dark matter components \citep[e.g.][]{2014ApJ...788L..35S, 2019MNRAS.489.2049N}. In these models, the baryonic mass is assumed to trace the light of the lens galaxy, often described by combinations of Sérsic profiles or Multi-Gaussian Expansions (MGEs) \citep{2002MNRAS.333..400C, 2024MNRAS.532.2441H}, with the mass-to-light ratio a free parameter. These methodological advances have been crucial in reducing false positives and improving the robustness of dark matter subhalo detections. 

Systematic searches for dark matter subhalos have so far relied on HST imaging, such as from the SLACS (Sloan Lens ACS Survey) and BELLS (BOSS Emission-Line Lens Survey) surveys \citep[][]{2014MNRAS.442.2017V, 2019MNRAS.485.2179R, 2024MNRAS.52710480N}. However, the limited angular resolution of HST ($\gtrsim$0.1 arcsec) constrains the sensitivity to subhalos of mass $M_{200} \gtrsim 10^9 \, \Msol$ \citep{2022MNRAS.510.2480D, 2022MNRAS.512.5862H}. For these masses, CDM and WDM predict nearly identical subhalo abundances. Additional challenges include contamination from imperfect subtraction of the lens galaxy’s light, which can mimic or obscure genuine small-scale perturbations \citep{2024MNRAS.532.2441H}, and the low redshift of most HST background sources ($z \sim 2$), which limits the cosmological volume probed. Many of these issues are likely to persist even in next-generation optical and near-infrared surveys with JWST \citep[e.g.][]{2025MNRAS.539..704L} and Euclid \citep[e.g.][]{2023MNRAS.521.2342O, 2025arXiv250116139W}.

ALMA offers a powerful alternative for probing dark matter substructure. Its long-baseline configurations can achieve angular resolutions as fine as 10 mas, over an order of magnitude improvement over HST, allowing sensitivity to subhalos down to $M_{200} \sim 10^7$–$10^8 \, \Msol$ \citep[e.g.][]{2016ApJ...823...37H, 2025A&A...693L..17S, 2025arXiv250802776S, 2025MNRAS.540L..78D}. Crucially, ALMA has now accumulated a substantial sample of strongly lensed dusty star-forming galaxies (DSFGs) observed at $\lesssim$100 mas resolution. This marks the first time that systematic, statistically meaningful searches for dark matter subhalos can be carried out using submm/mm-wave data. The compact, clumpy dust emission from DSFGs makes them highly sensitive to small-scale lensing perturbations, and the absence of foreground light removes a major systematic effect that afflicts optical/near-infrared analyses.

In this work, we present a detailed strong lensing analysis of ALMA data for PJ011646, with the aim of testing for the presence of a dark matter subhalo. The lens lies at  redshift $z=0.555$ and the source at  $z=2.125$ \citep[][]{2024ApJ...961....2K}.  In Section~\ref{sec:section_2}, we introduce the ALMA observations and describe the data reduction and post-processing procedures. In Section~\ref{sec:section_3}, we detail the lens modelling methodology, including the parameterization of the macro and subhalo models. In Section~\ref{sec:section_4}, we present the results for both the macro-model and subhalo search. In Section~\ref{sec:section_5}, we discuss the interpretation of the subhalo detection, evaluate the sensitivity of the data, and place constraints on the minimum detectable subhalo mass. Finally, in Section~\ref{sec:section_6}, we summarise our main conclusions and discuss the implications of our results for future dark matter subhalo studies with ALMA. Throughout this work, we adopt a spatially-flat $\Lambda$CDM cosmology with H$_0=67.8 \pm 0.9$\,km s$^{-1}$ Mpc$^{-1}$ and $\Omega_{\rm m}=0.308 \pm 0.012$ \citep{2016A&A...594A..13P}.

%%%%%%%%%%%%%%%%%%%%%%%%%%%%%%%%%%%%%%%%%%%%%%%%%%%%%%%%%%%%%%%%%%%%%%%%%%%
% SECTION
%%%%%%%%%%%%%%%%%%%%%%%%%%%%%%%%%%%%%%%%%%%%%%%%%%%%%%%%%%%%%%%%%%%%%%%%%%%
\section{Data} \label{sec:section_2}

The strongly lensed galaxy we analyze in this work, PJ011646, was observed with ALMA as part of the project,  2022.1.01311.S (P.I.: P.\ Kamieneski), in Band 7. The data were taken in two execution blocks with a total integration time of $\sim$10 mins. 

The data were calibrated using the standard ALMA pipeline in the Common Astronomy Software Applications package \citep[CASA;][]{2007ASPC..376..127M}. For visualization, we produce a cleaned image, using \textit{natural} weighting, of the dust continuum which is shown in the left panel of Figure~\ref{fig:data}. The cleaned image yields a synthesized beam of $0.096 \times 0.115$ arcsec and position angle of -71 deg. The root mean square (RMS) of the data reaches $\sim$0.06 mJy beam$^{-1}$ estimated around the lensed emission (i.e. background). No emission lines were detected in any of the spectral windows and, therefore, all are used in our analysis. In the right panel of Figure~\ref{fig:data}, we show HST/WFC3 data of the same lens in the F160W filter (P.I.: J.\ Lowenthal). These data are included for visualisation purposes only and are not used in the subsequent analysis.

Before we perform our lens modelling analysis, we execute the following data reduction steps. Each of the four spectral windows has 128 individual channels in a $\sim$2 GHz bandwidth. We average the visibilities from all channels into a single complex number, corresponding to the average point in the uv-plane (no additional time averaging was performed). Following the frequency averaging step, we calculate the noise of the data undertaking a similar approach to the one described in \cite{2021MNRAS.501..515P, 2022MNRAS.516.1808P}. First, we split visibilities by baseline, spectral window and polarization. Then, we subtract time-adjacent visibilities from one another which aims to remove the sky signal. Finally, we take the RMS of these visibilities (corrected by $\sqrt{2}$ to account for the subtraction) and assign it as the noise for this partition of the data. The noise estimation was performed separately for the two execution blocks as these were taken at different times.

%%%%%%%%%%%%%%%%%%%%%%%%%%%%%%%%%%%%%%%%%%%%%%%%%%%%%%%%%%%%%%%%%%%%%%%%%%%
% SECTION
%%%%%%%%%%%%%%%%%%%%%%%%%%%%%%%%%%%%%%%%%%%%%%%%%%%%%%%%%%%%%%%%%%%%%%%%%%%
\section{Methods} \label{sec:section_3}

All modelling presented in this work is performed using the open-source strong lensing software \textsc{PyAutoLens} \citep{2021JOSS....6.2825N}, together with its parent libraries. These include \textsc{PyAutoFit}, which provides a statistical framework for model fitting and Bayesian inference \citep{2021JOSS....6.2550N}, \textsc{PyAutoGalaxy}, which implements a comprehensive suite of mass profiles commonly used in lens modelling \citep{2023JOSS....8.4475N}, and \textsc{PyAutoArray}, which handles the underlying numerical operations. 

\textsc{PyAutoArray} provides efficient tools for converting image-plane surface-brightness distributions into interferometric visibilities through non-uniform fast Fourier transforms (NUFFT) and performs the associated linear algebra calculations. For interferometric datasets, these computations are optimized to exploit multiprocessing architectures, significantly improving performance.

For all modelling stages, we use the non-linear sampler \textsc{nautilus} \citep[][]{2023MNRAS.525.3181L}, which uses Importance Nested Sampling to build a Bayesian Posterior estimate.

\subsection{Mass models}

We model both the large-scale mass distribution of the lensing galaxy (i.e. the macro-model) and perturbative contributions from dark matter subhalos. Below we describe the parameterisations adopted for each component.

\subsubsection{Macro model}\label{sec:mass_model_macro}

The mass distribution of the lensing galaxy is represented by an elliptical power-law mass density profile \citep[EPL; ][]{2015A&A...580A..79T} with external shear. The convergence of the EPL profile is given by
\begin{equation}\label{eq:epl_mass}
    \kappa_{\rm EPL}(x, y) = \frac{3 - \alpha}{1 + q}
    \left( \frac{R_{\rm E}}{\sqrt{x^2 + y^2 / q^2}} \right)^{\alpha-1} \, ,
\end{equation}
where $\alpha$ is the three-dimensional logarithmic density slope, $q$ is the axis ratio (minor-to-major), and $R_{\rm E}$ is the Einstein radius. The profile is further characterised by a centroid position $(x_{\rm c}, y_{\rm c})$ and a position angle $\theta$, defined counterclockwise from the positive $x$-axis. Instead of fitting directly in terms of the axis ratio and position angle, we re-parametrise them in terms of two components given by,
\begin{equation}
    e_1 = \frac{1 - q}{1 + q} \sin\left(2\theta\right) \, , \qquad
    e_2 = \frac{1 - q}{1 + q} \cos\left(2\theta\right) \, ,
\end{equation}
which avoids degeneracies and discontinuities in parameter space \citep[e.g.][]{2022MNRAS.517.3275E, 2024MNRAS.52710480N}. 

Line-of-sight contributions are modelled with an external shear component. This shear is commonly parameterized by its magnitude, $\gamma_{\rm ext}$, and orientation, $\theta_{\rm ext}$. However, following the same convention as for the EPL mass distribution, we express it in terms of two components, $\gamma_{\rm ext,1}$ and $\gamma_{\rm ext,2}$, defined by
\begin{equation}
    \gamma_{\rm ext} = \sqrt{\gamma^2_{\rm ext, 1} + \gamma^2_{\rm ext, 2}} \, , \qquad
    \tan(2\theta_{\rm ext}) = \frac{\gamma_{\rm ext, 2}}{\gamma_{\rm ext, 1}} \, .
\end{equation}

To capture deviations from elliptical symmetry we also include higher-order angular perturbations in the form of Fourier multipoles \citep[e.g.][]{2016ApJ...823...37H, 2021MNRAS.507.1662M, 2022MNRAS.516.1808P, 2022A&A...663A.179V, 2024MNRAS.528.7564B, 2024MNRAS.52710480N, 2024MNRAS.528.1757O, 2024A&A...688A.110S}. The convergence of a multipole of order $m$ is
\begin{equation}
    \kappa_{\rm MP}(r, \phi) = \frac{1}{2} 
    \left(\frac{R_{\rm E}}{r} \right)^{\alpha - 1}
    k_m \cos\left[m (\phi - \phi_m)\right] ,
\end{equation}
where $k_m$ and $\phi_m$ denote the amplitude and phase of the perturbation, respectively. The quantities $R_{\rm E}$ and $\alpha$ are fixed to be identical to those of the EPL component. The corresponding deflection field, obtained by solving the Poisson equation in polar coordinates, is
\begin{align}
    a_r(r, \phi) &= - \frac{3 - \alpha}{m^2 - (3 - \alpha)}
    R_{\rm E}^{\alpha - 1} r^{2 - \alpha} k_m
    \cos\left[m(\phi - \phi_m)\right] , \\
    a_{\phi}(r, \phi) &= \frac{m^2}{m^2 - (3 - \alpha)}
    R_{\rm E}^{\alpha - 1} r^{2 - \alpha} k_m
    \sin\left[m (\phi - \phi_m)\right] .
\end{align}
These are converted to Cartesian deflections via standard coordinate transformations. As with the parameters of the EPL and shear, we parameterise $(k_m, \phi_m)$ using two components
\begin{equation}
    e^{m}_{1} = k_m \cos(m\phi_m) \, , \qquad
    e^{m}_{2} = k_m \sin(m\phi_m) \, ,
\end{equation}
which avoids discontinuities and degeneracies associated with the periodic nature of $\phi_m$, and provides a smooth, well-behaved parameter space for all values of $m$ (e.g. $m = 4$ is invariant under $45^\circ$ rotations).

\subsubsection{Subhalos}\label{sec:subhalo_model}

We model dark matter subhalos with a spherical Navarro-Frenk-White (NFW) profile. This profile represents a universal description for the density distribution of dark matter (sub-)halos in N-body cosmological simulations \citep{1996ApJ...462..563N, 1997ApJ...490..493N}. Its volume mass density is given by,
\begin{equation}
    \rho(r) = \frac{\rho_{\rm s}}{(r/r_{\rm s}) (1 + r/r_{\rm s})^2} \, ,
    \label{eqn:subhalo_1}
\end{equation}
where $\rho_s$ is the density normalization and $r_s$ the scale radius (the radius where the logarithmic slope is -2). The convergence is given by,
\begin{equation}
    \kappa(R) = 2 \kappa_s g\left( \frac{R}{r_s} \right) \, ,
    \label{eqn:subhalo_2}
\end{equation}
where the dimensionless function, $g(x)$, is given by \citep[][]{1996A&A...313..697B},
\begin{equation}
    g(x) =
    \begin{cases}
        \dfrac{1}{x^2 - 1} \left[ 1 - \dfrac{2}{\sqrt{1 - x^2}}
        \operatorname{arctanh} \!\left( \sqrt{\dfrac{1 - x}{1 + x}} \right) \right],
        & 0 \le x < 1, \\[12pt]
        \dfrac{1}{3}, & x = 1, \\[8pt]
        \dfrac{1}{x^2 - 1} \left[ 1 - \dfrac{2}{\sqrt{x^2 - 1}}
        \arctan \!\left( \sqrt{\dfrac{x - 1}{1 + x}} \right) \right],
        & x > 1.
    \end{cases}
\end{equation}
and $\kappa_s = \rho_s r_s / \Sigma_{\rm crit}$, where $\Sigma_{\rm crit}$ is the critical surface density that depends on the lens and source redshifts.

As a generalization, we also consider the generalized NFW \citep[gNFW][]{1996MNRAS.278..488Z} profile, which introduces an inner logarithmic slope $\gamma$ as a free parameter:
\begin{equation}
\rho(r) = \frac{\rho_{\rm s}}{(r/r_{\rm s})^{\gamma}  (1 + r/r_{\rm s})^{3 - \gamma}} \, .
\end{equation}
The standard NFW form is recovered for $\gamma = 1$, while deviations from this value allow for either shallower or steeper inner density slopes. This added flexibility enables tests of whether the inferred subhalo profiles are consistent with expectations from CDM simulations or suggest an alternative inner structure \citep[e.g.][]{2025arXiv251017956K}.

\subsection{Source inversion}\label{sec:source_inversion}

We reconstruct the source surface brightness distribution following the Bayesian formalism for pixelated sources introduced by \citet{2006MNRAS.371..983S}. In this framework, we infer the most probable source pixel values, $\mathbf{s}$, given the interferometric data (visibilities) $\mathbf{d}$, a fixed lens model $\boldsymbol{\theta}$, and a set of source hyper-parameters $\boldsymbol{\lambda}$. The posterior probability is written as
\begin{equation}\label{eq:likelihood}
\begin{aligned}
-2\log P(\mathbf{d} | \boldsymbol{\theta}, \boldsymbol{\lambda}) = \, & \chi^2 + s^T \mathbf{H} s + \log\det(\mathbf{F} + \mathbf{H}) - \log\det\mathbf{H} \\
&- \log\det(2\pi\mathbf{C}^{-1}) 
\end{aligned}
\end{equation}
In the above equation, the term, $\mathbf{H}$, is the regularization matrix and depends on $\lambda$ \citep[][]{2015MNRAS.452.2940N}. The $\chi^2$ term is given by
\begin{equation}\label{eq:chi_square}
\chi^2 = \left( \mathbf{D} \mathbf{L} s - d\right)^T \mathbf{C}^{-1} \left( \mathbf{D} \mathbf{L} s - \mathbf{d}\right) \, ,
\end{equation}
where $\mathbf{L}$ is the matrix mapping image-plane pixels to source-plane pixels which depends on $\theta$, and $\mathbf{D}$ is a matrix that converts the sky surface brightness into complex visibilities (normally a non-uniform discrete Fourier transform). Finally, the matrix $\mathbf{F}$, referred to as the curvature matrix in \cite{2015MNRAS.452.2940N}, is given by $\rm \mathbf{F} = \mathbf{L}^T \mathbf{D}^T \mathbf{C}^{-1} \mathbf{D} \mathbf{L}$.

Maximizing $P(s | \mathbf{d}, \boldsymbol{\theta}, \boldsymbol{\lambda})$ for fixed $\boldsymbol{\theta}$ and $\boldsymbol{\lambda}$ yields the regularized least-squares equation,
\begin{equation}
\left( \mathbf{L}^T \mathbf{D}^T \mathbf{C}^{-1} \mathbf{D} \mathbf{L} + \mathbf{H} \right) s_{\rm MP} = \mathbf{L}^T \mathbf{D}^T \mathbf{C}^{-1} \mathbf{d} \, ,
\end{equation}
where $s_{\rm MP}$ denotes the maximum a posteriori (MAP) source.

The term $\mathbf{D}^T \mathbf{C}^{-1} \mathbf{D}$, which we denote as $\tilde{\mathbf{C}}^{-1}$, is the image-plane noise covariance matrix. This operation is computationally expensive and would need to be repeated for every lens model evaluation if computed directly. However, because it is independent of the lens parameters $\boldsymbol{\theta}$, we compute it once at the start of the analysis using fast Fourier transforms (FFTs). Likewise, we precompute the term $\mathbf{D}^T \mathbf{C}^{-1} \mathbf{d}$, denoted as $\tilde{\mathbf{d}}$, which corresponds to the dirty image.

By precomputing these two quantities, the inversion step becomes independent of the number of visibilities, as no FFTs are required to evaluate the likelihood for new sets of lens parameters. The only FFT performed during the modelling process is that used to transform a model lensed image into visibilities for the computation of the $\chi^2$ term (Eq.~\ref{eq:chi_square}).  The most computationally expensive part of the likelihood evaluation is the matrix product $\mathbf{L}^T \tilde{\mathbf{C}}^{-1} \mathbf{L}$, which scales with the number of image- and source-plane pixels. We exploit symmetries in the $\tilde{\mathbf{C}}^{-1}$ matrix to parallelize this computation, significantly reducing the overall runtime\footnote{The details of this implementation are available at \url{https://github.com/PyAutoLabs/PyAutoLens}}.

Finally, in order to reconstruct the lensed surface brightness distribution we use a Voronoi mesh pixelization and adopt a natural neighbour interpolation scheme \citep[e.g.][]{2024MNRAS.532.2441H}. The strength of regularization adapts to the surface brightness of the reconstructed source \citep[e.g.][]{2015MNRAS.452.2940N} and uses interpolated values in a cross pattern for each traced image-plane pixel to estimate the regularization matrix. The details of this method are described in Appendix~A of \citet{2024MNRAS.532.2441H}.

%%%%%%%%%%%%%%%%%%%%%%%%%%%%%%%%%%%%%%%%%%%%%%%%%%%%%%%%%%%%%%%%%%%%%%%%%%%
% SECTION
%%%%%%%%%%%%%%%%%%%%%%%%%%%%%%%%%%%%%%%%%%%%%%%%%%%%%%%%%%%%%%%%%%%%%%%%%%%
\section{Results} \label{sec:section_4}

\begin{table}
\caption{Posterior distributions of the model parameters. The reported values correspond to the median of the marginalised posterior distributions, with 1$\sigma$ uncertainties indicating the 16th–84th percentile ranges. The log Bayesian evidence difference, $\Delta \ln Z$, is given relative to the PL model.}
\renewcommand{\arraystretch}{1.5}
\centering

\begin{tabular}{crrr}
& PL & PL + MP & PL + MP \\
& & $(m = 3, 4)$ & $(m = 1, 3, 4)$ \\
\hline
$R_{\rm E}$ 
& $2.225_{-0.008}^{+0.011}$ 
& $2.215_{-0.011}^{+0.009}$ 
& $2.217_{-0.015}^{+0.021}$ \\

$e_{1}$ 
& $-0.207_{-0.002}^{+0.002}$ 
& $-0.176_{-0.004}^{+0.005}$ 
& $-0.179_{-0.006}^{+0.004}$ \\

$e_{2}$ 
& $0.171_{-0.003}^{+0.003}$ 
& $0.162_{-0.003}^{+0.004}$ 
& $0.182_{-0.005}^{+0.005}$ \\

$\alpha$ 
& $1.626_{-0.006}^{+0.009}$ 
& $1.627_{-0.009}^{+0.009}$ 
& $1.632_{-0.012}^{+0.016}$ \\

$e_{1}^{\rm ext}$ 
& $0.070_{-0.002}^{+0.002}$ 
& $0.065_{-0.002}^{+0.003}$ 
& $0.080_{-0.003}^{+0.003}$ \\

$e_{2}^{\rm ext}$ 
& $-0.118_{-0.001}^{+0.001}$ 
& $-0.096_{-0.002}^{+0.003}$ 
& $-0.098_{-0.003}^{+0.004}$ \\

$e_{1}^{m_{1}}$ 
& - 
& - 
& $0.034_{-0.005}^{+0.004}$ \\

$e_{2}^{m_{1}}$ 
& - 
& - 
& $-0.002_{-0.011}^{+0.009}$ \\

$e_{1}^{m_{3}}$ 
& - 
& $0.013_{-0.001}^{+0.002}$ 
& $0.017_{-0.002}^{+0.002}$ \\

$e_{2}^{m_{3}}$ 
& - 
& $0.008_{-0.001}^{+0.002}$ 
& $0.011_{-0.001}^{+0.001}$ \\

$e_{1}^{m_{4}}$ 
& - 
& $-0.008_{-0.002}^{+0.002}$ 
& $-0.009_{-0.002}^{+0.002}$ \\

$e_{2}^{m_{4}}$ 
& - 
& $-0.012_{-0.002}^{+0.002}$ 
& $-0.013_{-0.002}^{+0.002}$ \\
\hline
$\Delta \ln Z$ & 0 & 52.1 & 58.6 \\
\end{tabular}
\label{tab:parameters}
\end{table}

\subsection{Macro-model}

\begin{figure}
\centering
\begin{tabular}{c}
\includegraphics[width=0.95\columnwidth]{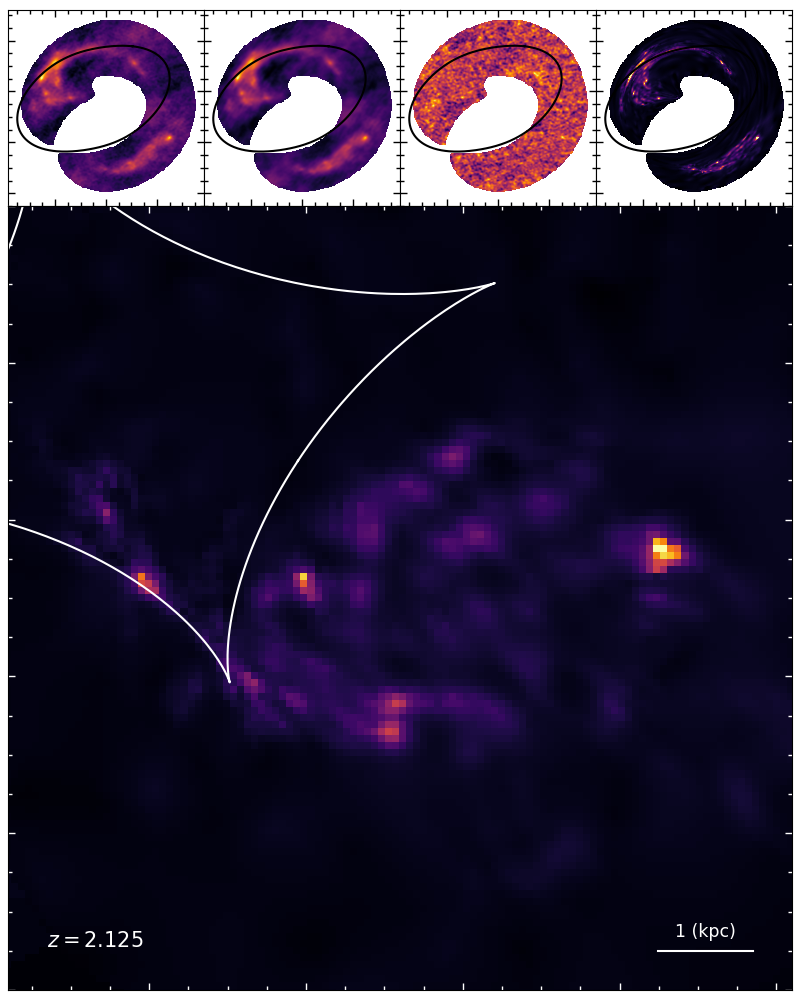} \\
\includegraphics[width=0.95\columnwidth]{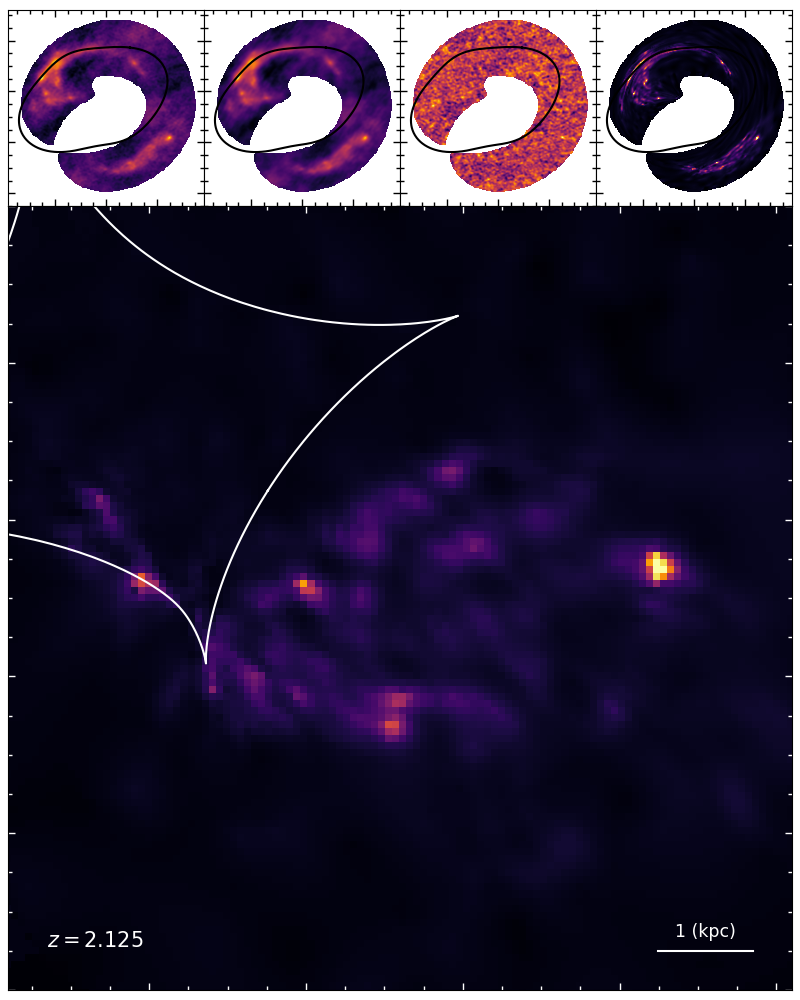}  \\
\end{tabular}
\caption{Model source surface brightness distributions obtained using the PL (top) and PL+MP (bottom) macro-models. The top panels show, from left to right, the data, model, residuals and model lensed surface brightness distribution. The black line in the top panels corresponds to the critical curve, while the white lines in the bottom panels correspond to the caustic curves. The residuals are plotted on the same range for both models.}
\label{fig:reconstructions}
\end{figure}

We first compare two parameterizations of the lens macro-model: an elliptical power-law (PL) profile with external shear, and a more flexible model that includes additional multipole terms (PL+MP) of orders $m = 3$ and $m = 4$ \citep[e.g.][]{2022MNRAS.516.1808P, 2024A&A...688A.110S}. The source reconstructions obtained for these two models are shown in Figure~\ref{fig:reconstructions}, and the corresponding posterior values of the model parameters are reported in Table~\ref{tab:parameters}. 

%%%%%%%%%%%%%%%%%%%%%%%%%%%%%%%%%%%%%%%%%%%%%%%%%%%%%%%%%%%%%%%%%%%%%%%%%%%%%%%%
% FIGURE
%%%%%%%%%%%%%%%%%%%%%%%%%%%%%%%%%%%%%%%%%%%%%%%%%%%%%%%%%%%%%%%%%%%%%%%%%%%%%%%%
\begin{figure*}
\centering
\begin{tabular}{cc}
\includegraphics[width=0.5\textwidth]{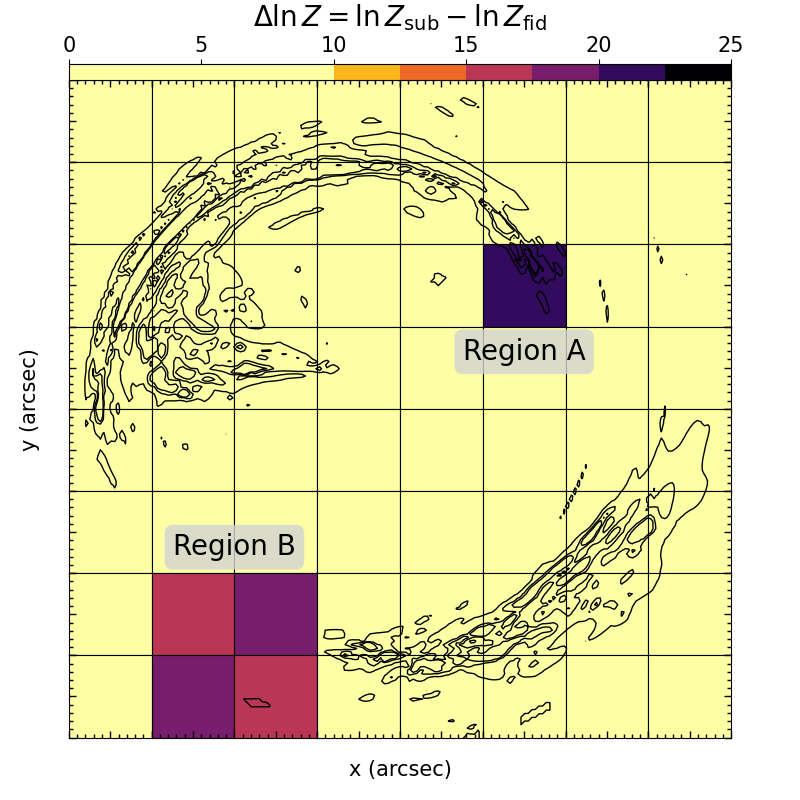} \hspace{-10mm}
& \includegraphics[width=0.5\textwidth]{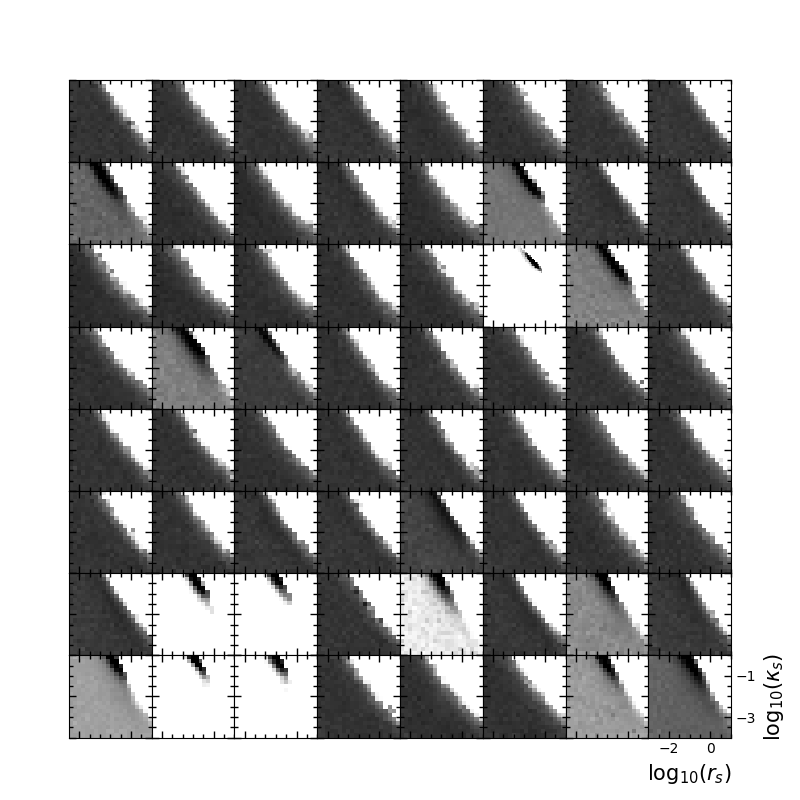}  \\
\end{tabular}
\caption{\textit{(Left):} Log evidence difference map of a model with versus a model without a dark matter subhalo, $\Delta \ln Z = \ln Z_{\rm sub} - \ln Z_{\rm fid}$. The black contours correspond to the observed surface brightness distribution of the lensed source. \textit{(Right):} Posterior probability distributions for the subhalo parameters ($r_s$, $\kappa_s$) for each of the grid cells.}
\label{fig:subhalo_[1]}
\end{figure*}
%%%%%%%%%%%%%%%%%%%%%%%%%%%%%%%%%%%%%%%%%%%%%%%%%%%%%%%%%%%%%%%%%%%%%%%%%%%%%%%%
% END
%%%%%%%%%%%%%%%%%%%%%%%%%%%%%%%%%%%%%%%%%%%%%%%%%%%%%%%%%%%%%%%%%%%%%%%%%%%%%%%%

The PL+MP model is strongly preferred, with a difference in Bayesian log-evidence of $\Delta \ln Z = 57.4$ relative to the PL model. Consistent with the findings of \citet{2022MNRAS.516.1808P} and \citet{2024A&A...688A.110S}, this improvement in evidence is primarily driven by a change in the source regularization terms rather than by a substantial reduction in the uv-plane residuals. The change in the $\chi^2$ term of the log-likelihood function (Eq.~\ref{eq:likelihood}) is modest ($\Delta \chi^2 \approx 10$), indicating that both models reproduce the observed data comparably well (in uv-space). The marginalized posterior distributions for the outer regularization coefficient (see Section~\ref{sec:source_inversion}) are consistent between the PL and PL+MP models, however, the median value is slightly higher for the PL+MP case. This reflects the additional flexibility of the multipole terms, which allow the source to be more smoothly reconstructed. The improvement is most visually evident in the part of the source lying within the caustic curve (see Figure~\ref{fig:reconstructions}), which appears more sharply focused in the PL+MP reconstruction.

The parameters of the PL component are generally consistent between the two models, with only minor differences in the inferred axis ratio and the magnitude of the external shear. The best-fitting amplitudes of the third- and fourth-order multipoles are $k_3 = 1.54_{-0.17}^{+0.21}$ and $k_4 = 1.47_{-0.23}^{+0.24}$ per cent of the convergence, respectively. These values are typical for galaxy-scale lenses and consistent with expectations from previous analyses \citep[e.g.][]{2022MNRAS.516.1808P, 2024A&A...688A.110S, 2025MNRAS.540..247E, 2025MNRAS.543..540T}. Recent work by \citet{2025PhRvD.111l3014P} has shown that the commonly used spherical multipole formalism can introduce inaccuracies when applied to highly flattened lenses. Since the lens galaxy in PJ011646 has an inferred axis ratio of $q \simeq 0.6$, this approximation may lead to mild biases in the recovered multipole amplitudes. An alternative elliptical multipole model has been proposed to mitigate this issue; however, it is currently limited to SIE profiles. As our best-fitting model favours a slope significantly shallower than isothermal, we do not adopt this approach in our analysis.

We also explored a model including a first-order multipole ($m = 1$) in addition to the $m = 3$ and $m = 4$ components. The resulting Bayesian log-evidence is $\Delta \log Z = 58.6$, only 6.5 higher than that of the fiducial PL+MP model, suggesting that the inclusion of the dipole term provides only a marginal improvement. In this case, the amplitudes of the multipoles are $k_1 = 3.4_{-0.2}^{+0.4}$, $k_3 = 2.02_{-0.20}^{+0.21}$, and $k_4 = 1.51_{-0.30}^{+0.26}$ per cent of the convergence. \citet{2025MNRAS.540.3281A} found that first-order multipole amplitudes above $\sim$2–3 per cent are typically associated with galaxies experiencing a recent or ongoing interaction. In the case of PJ011646, nearby galaxies with similar photometric redshifts lie at projected distances of $\sim$50~kpc ($\sim$8 arcsec away), making it possible that the measured $m=1$ term is driven by a physical companion.

Finally, this system was previously modelled by \citet{2024ApJ...961....2K} and \citet{2024NatAs...8.1181L}, although neither study reported the best-fitting parameters of their adopted macro-models. Both analyses were based on different datasets than those used here—specifically, lower-resolution ALMA Band 7 observations and HST imaging in the F160W filter, respectively. Consequently, a direct comparison with our results is not possible.

\subsection{Subhalo search}

\subsubsection{Initial search}

Next, we search for dark matter subhalos by adding an NFW mass distribution (with log-uniform priors on parameters $-4 < \log_{10}(\kappa_{\rm s}) < 0$ and $-3 < \log_{10}(r_{\rm s}/{\rm kpc}) < 1$; see Section~\ref{sec:subhalo_model}), while re-optimizing all macro-model parameters and source hyper-parameters. To facilitate optimisation of the subhalo position $(x_{\rm sub}, y_{\rm sub})$, we follow the procedure of \citet{2024MNRAS.52710480N, 2025MNRAS.539..704L} and search with a uniform prior inside each cell of an $8\times 8$ grid. This produces a map of the Bayesian log-evidence difference, $\Delta \ln Z$, between models with and without a subhalo (Figure~\ref{fig:subhalo_[1]}). We adopt a threshold $\Delta \ln Z > 10$ to consider a robust preference for the model with a subhalo. This corresponds approximately to a $\gtrsim 3\sigma$ detection\footnote{
To convert a difference in log Bayesian evidence, $\Delta \ln Z$, into an approximate Gaussian significance, we assume $\Delta \chi^2 \approx 2 \Delta \ln Z$, which is valid under the Laplace approximation for nested models with approximately Gaussian likelihoods.  
We then compute a $p$-value using the $\chi^2$ cumulative distribution function:
\begin{equation*}
    p = 1 - F(\Delta \chi^2, \nu),
\end{equation*}
where $\nu$ is the number of additional parameters in the more complex model and $F$ is the cumulative $\chi^2$ distribution.  
Finally, we convert $p$ to an approximate Gaussian ``$\sigma$'' via, $\sigma = \Phi^{-1}(1 - p)$, where $\Phi^{-1}$ is the inverse standard normal cumulative distribution function.
}.

Two distinct regions of enhanced evidence for subhalos are identified, where $\Delta \ln Z > 10$: (i) a single cell located right of the northern arc, with $\Delta \ln Z \approx 22$, and (ii) a cluster of four neighbouring cells left of the southern arc, with a maximum $\Delta \ln Z \approx 18$. Black contours in Figure~\ref{fig:subhalo_[1]} show the observed surface brightness distribution of the lensed emission. Across the remainder of the grid, the log-evidence difference is less than 10. These cells are consistent with the absence of a subhalo, and also provide meaningful upper limits on the masses of any subhalos that could reside there (see Section \ref{sec:section_5_1}).

In the right panel of Figure~\ref{fig:subhalo_[1]} we present the posterior two-dimensional distributions of the subhalo parameters, $\kappa_s$ and $r_s$, for each individual grid cell in the map shown in left panel of the same figure. The spatial extent of each panel corresponds to the log-uniform prior limits adopted for these parameters. Cells with a Bayesian evidence difference $\Delta \ln Z > 10$ are highlighted in purple, whereas the remaining cells are shown in grey.

%%%%%%%%%%%%%%%%%%%%%%%%%%%%%%%%%%%%%%%%%%%%%%%%%%%%%%%%%%%%%%%%%%%%%%%%%%%%%%%%
% FIGURE
%%%%%%%%%%%%%%%%%%%%%%%%%%%%%%%%%%%%%%%%%%%%%%%%%%%%%%%%%%%%%%%%%%%%%%%%%%%%%%%%
\begin{figure}
\centering
\includegraphics[width=0.995\columnwidth]{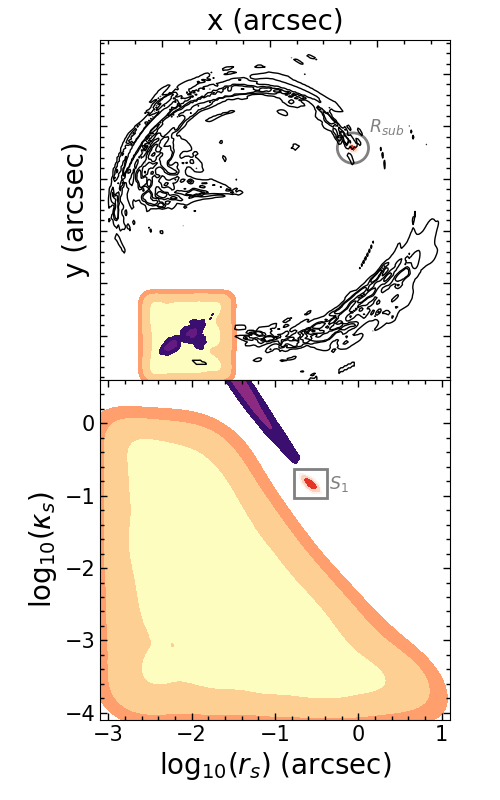} 
\caption{Posterior probability distributions for the parameters of the trial subhalo located left of the southern arc, shown for both the one-subhalo (purple) and two-subhalo (orange) model fits. The corresponding posteriors for the subhalo located to the right of the northern arc in the two-subhalo model are also shown in red. {\it (Top panel)}: $x_{\rm sub}$ and $y_{\rm sub}$; {\it (Bottom panel)}: $\kappa_s$ and $r_s$.}
\label{fig:temp_2}
\end{figure}
%%%%%%%%%%%%%%%%%%%%%%%%%%%%%%%%%%%%%%%%%%%%%%%%%%%%%%%%%%%%%%%%%%%%%%%%%%%%%%%%
% END
%%%%%%%%%%%%%%%%%%%%%%%%%%%%%%%%%%%%%%%%%%%%%%%%%%%%%%%%%%%%%%%%%%%%%%%%%%%%%%%%

%%%%%%%%%%%%%%%%%%%%%%%%%%%%%%%%%%%%%%%%%%%%%%%%%%%%%%%%%%%%%%%%%%%%%%%%%%%%%%%%
% FIGURE
%%%%%%%%%%%%%%%%%%%%%%%%%%%%%%%%%%%%%%%%%%%%%%%%%%%%%%%%%%%%%%%%%%%%%%%%%%%%%%%%
\begin{figure*}
\centering
\includegraphics[width=0.995\textwidth]{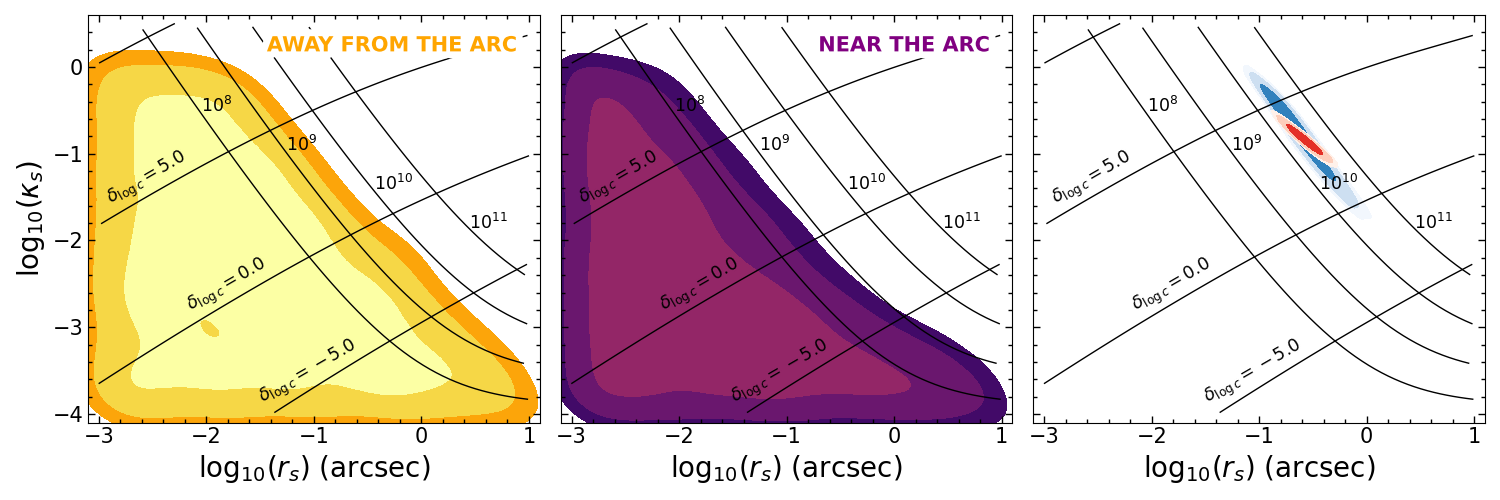} 
\caption{Posterior probability distributions for the subhalo parameters ($r_s$ and $\kappa_s$) at various locations across the image plane: away from the lensed arcs (left panel), near the northern lensed arc (middle panel), and in the cell containing the detected subhalo (right panel), shown for both the NFW (red) and gNFW (blue) models. The black lines correspond to tracks in this 2D plane of constant $M_{200}$ and concentration offset, $\delta_{\log c}$, assuming the \protect\citet{2016MNRAS.460.1214L} mass--concentration relation.}
\label{fig:temp_1}
\end{figure*}
%%%%%%%%%%%%%%%%%%%%%%%%%%%%%%%%%%%%%%%%%%%%%%%%%%%%%%%%%%%%%%%%%%%%%%%%%%%%%%%%
% END
%%%%%%%%%%%%%%%%%%%%%%%%%%%%%%%%%%%%%%%%%%%%%%%%%%%%%%%%%%%%%%%%%%%%%%%%%%%%%%%%

\subsubsection{Interpretation of the subhalo map}

For the two regions with $\Delta \ln Z > 10$, two interpretations are possible: (i) both locations correspond to genuine subhalos, or (ii) only one contains a real subhalo, while the other is a spurious preference produced by model degeneracies. In particular, when two different image-plane regions map to the same part of the source plane, then a mis-modelled perturbation in one location can manifest as an apparent improvement when a subhalo is placed in the other. We discuss this degeneracy further below. To distinguish between these scenarios, we perform an additional fit with a model that includes two subhalos. For the first subhalo (right of the northern arc), we adopt Gaussian priors centred on the posterior distributions obtained from the grid search. For the second subhalo, we use the same uniform priors as in the initial search for $\kappa_{\rm s}$ and $r_{\rm s}$, and restrict its position to within the four cells left of the southern arc. Because combining the posterior distributions from neighbouring cells can introduce boundary artefacts, we also perform an additional fit including a single subhalo confined within the combined region defined by these four cells in order to accurately estimate the posterior distribution of a subhalo in that region.

We find equal log-evidences when comparing the two-subhalo model to a single-subhalo model (right of the northern arc), implying that the data are consistent with only one subhalo being present. For the two-subhalo model, the posterior distribution of the position of the second subhalo (nominally located left of the southern arc) is effectively flat and indistinguishable from the prior, as shown in the top panel of Figure~\ref{fig:temp_2}. Likewise, the posterior distributions for its parameters, $\kappa_{\rm s}$ and $r_{\rm s}$, reduce to upper limits (bottom panel of Figure~\ref{fig:temp_2}), consistent with those of the neighbouring grid cells where the initial subhalo search yielded $\Delta \ln Z < 10$.

We also highlight that the two regions of high $\Delta \ln Z$ in the image plane correspond to similar regions in the source plane, traced using the best-fit macromodel. Consequently, if a real subhalo is located right of the northern arc (region A), omitting it from the model results in an incorrect reconstruction of the source structure in the corresponding area of the source plane. Because the southern-arc region (region B) maps to the same part of the source, the model will also perform poorly there. In this situation, introducing a modelled subhalo near region B can partially compensate for the missing perturbation at region A, artificially boosting the evidence at B despite no genuine subhalo being present. This behaviour is well documented in previous grid-based subhalo searches \citep[e.g.][]{2023MNRAS.518..220H, 2024MNRAS.52710480N} and is especially clear in mock tests containing only a single subhalo \citep[see Appendix~B of][]{2024MNRAS.52710480N}. \textcolor{black}{We note that an alternative to the grid-based approach is the potential-correction method \citep[e.g.][]{2010MNRAS.408.1969V, 2012Natur.481..341V, 2022MNRAS.516.1347V, 2025arXiv250419177C}, which adjusts the potential coherently across the image plane and therefore does not generate the bimodal $\Delta \ln Z$ features seen in subhalo maps like ours. However, this method offers a less direct connection between the inferred perturbation and a physically parameterized subhalo.}

Finally, support for the subhalo being located right of the northern arc comes from the physical properties inferred for the alternative (southern) candidate region. \textcolor{black}{The corresponding posterior distributions imply extremely high concentrations ($c_{200} > 1000$), which exceed the mass–concentration relation of \citet{2016MNRAS.460.1214L} by more than $30\sigma$ and are therefore unphysical in the context of $\Lambda$CDM.} We thus interpret the signal left of the southern arc as a false positive.

\subsubsection{Alternative subhalo models}

We additionally tested a gNFW profile for the subhalo in the grid cell corresponding to the highest Bayesian evidence difference. The resulting model yields a log Bayesian evidence equivalent to that of the NFW profile, indicating that the data do not require a deviation from the standard NFW form within the range of scales probed. The same log-uniform priors were adopted for $\kappa_{\rm s}$ and $r_{\rm s}$, while the inner density slope was allowed to vary uniformly within $0 < \gamma_{\rm gNFW} < 2$. From the posterior distribution, we obtain a $3 \sigma$ upper limit of $\gamma_{\rm gNFW} < 1.7$.

%%%%%%%%%%%%%%%%%%%%%%%%%%%%%%%%%%%%%%%%%%%%%%%%%%%%%%%%%%%%%%%%%%%%%%%%%%%
% SECTION
%%%%%%%%%%%%%%%%%%%%%%%%%%%%%%%%%%%%%%%%%%%%%%%%%%%%%%%%%%%%%%%%%%%%%%%%%%%
\section{Discussion} \label{sec:section_5}

In this section, we place our results in the broader context of dark matter substructure studies, comparing the inferred properties of the detected subhalo with previous detections in the literature. We also assess the sensitivity of the dataset, quantifying the minimum subhalo mass that could be detected with these observations.

\subsection{Upper limits from subhalo search}\label{sec:section_5_1}

We assess the sensitivity of our data to dark matter subhalos using the grid cells from the subhalo search with $\Delta \ln Z < 10$, for which the data do not favour the inclusion of a subhalo but can still provide meaningful upper limits on subhalo properties. In the left and middle panels of Figure~\ref{fig:temp_1}, we show the posterior distributions of the subhalo parameters $\kappa_{\rm s}$ and $r_{\rm s}$ for two representative grid cells: one located away from the lensed arcs (left panel) and one lying directly on top of the northern arc (middle panel). For both of these cells, the log-evidence difference relative to the macromodel-only fit is less than 10, indicating no significant preference for a subhalo. However, the posterior distributions constrain the allowed parameter space and can therefore be used to derive upper limits on the subhalo mass and concentration.

Our choice to parameterise subhalos in terms of $\kappa_{\rm s}$ and $r_{\rm s}$ allows for a direct comparison with theoretical predictions based on any assumed mass–concentration relation (MCR). In Figure~\ref{fig:temp_1}, we overlay tracks of constant halo mass, $M_{200}$, and concentration offset, $\delta_{\log c}$, derived using the \citet{2016MNRAS.460.1214L} MCR. The latter is expressed as a deviation from the median relation, parameterised by
\begin{equation}
    \log_{10} c_{200} = \log_{10} c_{200, \mathrm{MCR}} + \delta_{\log c} \,\sigma_{\log c} \, ,
\end{equation}
where $\sigma_{\log c} = 0.15 \, \mathrm{dex}$ quantifies the scatter in concentration at fixed mass, while $\delta_{\log c}$ describes how many standard deviations the inferred concentration lies above or below the MCR.

As shown in Figure~\ref{fig:temp_1}, cells located away from the lensed arcs allow a broad range of $(\kappa_{\rm s}, r_{\rm s})$ values, corresponding to relatively weak constraints on subhalo mass. The limits tighten considerably near the arcs, where the sensitivity to local perturbations in the lensing potential is highest. Along the bright northern arc, the posteriors exhibit well-defined upper bounds, excluding subhalos more massive than a few times $10^8$ to $10^9 \, \Msol$, depending on the assumed concentration. This behaviour is consistent with results from mock-lens sensitivity analyses \citep[e.g.][]{2019MNRAS.485.2179R, 2022MNRAS.510.2480D, 2023MNRAS.521.2342O}, which show that the detectability of subhalos is strongly correlated with the local surface brightness gradient and the magnification of the lensed image. 

\textcolor{black}{We note, however, that the upper bounds we derive here are not equivalent to those obtained from a sensitivity function analysis. In a sensitivity calculation, the intrinsic source surface brightness distribution is held fixed, and mock datasets are generated by inserting subhalos at different locations in the image-plane (the data that are being modelled change). One then asks what minimum subhalo mass produces a log-evidence difference above a chosen detection threshold between models with and without a subhalo \citep[e.g. $\Delta \ln Z > 50$;][]{2022MNRAS.510.2480D}. In our case, by contrast, the observed lensed surface brightness is fixed (i.e. the data), and we instead ask what is the maximum subhalo mass that can be added at a given location before it becomes inconsistent with the data at the $3\sigma$ level, roughly equivalent to $\Delta \ln Z > -10$.}

\textcolor{black}{Finally, in the ($\kappa_s$, $r_s$) plane, tracks of constant halo mass, $M_{200}$, follow an approximately negative gradient: increasing the scale radius requires a lower normalisation to preserve the total mass. The posterior constraints and upper limits derived from our grid cells mirror this trend. As shown in Figure~\ref{fig:temp_1}, the 1, 2 and $3\sigma$ contours form a roughly triangular region whose slope closely follows that of the constant mass tracks. This behaviour reflects the underlying degeneracy between $\kappa_{\rm s}$ and $r_{\rm s}$, the data primarily constrain the projected mass within the region of highest sensitivity, rather than the two parameters individually.}

%%%%%%%%%%%%%%%%%%%%%%%%%%%%%%%%%%%%%%%%%%%%%%%%%%%%%%%%%%%%%%%%%%%%%%%%%%%%%%%%
% FIGURE
%%%%%%%%%%%%%%%%%%%%%%%%%%%%%%%%%%%%%%%%%%%%%%%%%%%%%%%%%%%%%%%%%%%%%%%%%%%%%%%%
\begin{figure}
\centering
\includegraphics[width=0.9\columnwidth]{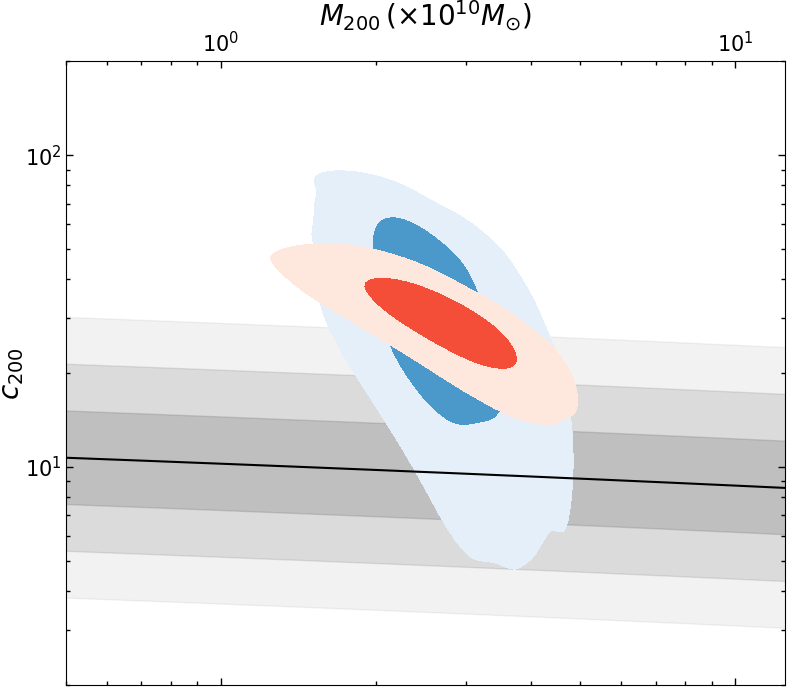} \\
\includegraphics[width=0.9\columnwidth]{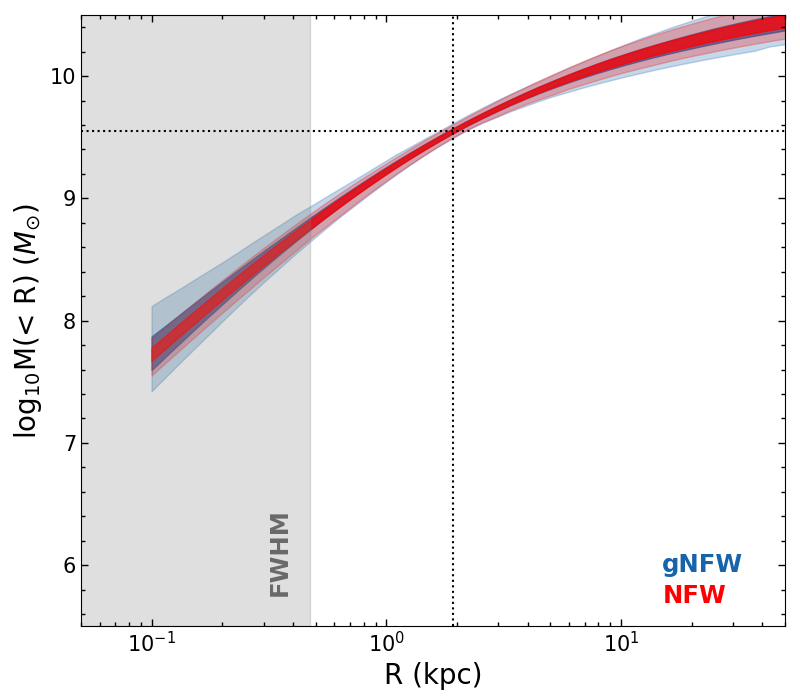}
\caption{\textit{(Top):} Posterior probability distribution for the parameters $M_{200}$ and $c_{200}$ for the NFW and gNFW profiles, shown in red and blue respectively. The black line corresponds to the mass-concentration relation by Ludlow et al. (2016) and the gray region to the $1\sigma$ scatter. \textit{(Bottom):}The projected enclosed mass profiles, $M( < R)$, corresponding to the best-fit subhalo parameters for the NFW and gNFW profiles, shown in red and blue respectively. The vertical dotted black line correspond to the distance where the error on the enclosed mass in the lowest for the NFW profile. The grey areas show the resolution of the data used to constrain these models.}
\label{fig:enclosed_mass}
\end{figure}
%%%%%%%%%%%%%%%%%%%%%%%%%%%%%%%%%%%%%%%%%%%%%%%%%%%%%%%%%%%%%%%%%%%%%%%%%%%%%%%%
% END
%%%%%%%%%%%%%%%%%%%%%%%%%%%%%%%%%%%%%%%%%%%%%%%%%%%%%%%%%%%%%%%%%%%%%%%%%%%%%%%%

\subsection{Properties of the detected subhalo}

In the right panel of Figure~\ref{fig:temp_1} we show posterior distributions for the subhalo parameters, $\kappa_s$ and $r_s$, from the grid cell with the highest log Bayesian evidence difference. Adopting the mass-concentration relation from \cite{2016MNRAS.460.1214L}, its concentration is, $\sigma_{\log c} = 3.2 \pm 0.5$, above the median concentration predicted by the relation.
% \begin{itemize}
%     \item 
% \end{itemize}

The subhalo parameters $(M_{200}, c_{200})$ can be derived from the lensing parameters $(\kappa_{\mathrm{s}}, r_{\mathrm{s}})$ by equating the NFW mass profile to the definition of $M_{200}$. The mass enclosed within $r_{200} = c_{200}\,r_{\mathrm{s}}$, where the mean density equals $200\,\rho_{\mathrm{crit}}(z_{\mathrm{l}})$, is
\begin{align}
M_{200} &= \frac{4\pi}{3}\,200\,\rho_{\mathrm{crit}}(z_{\mathrm{l}})\,r_{200}^3 \nonumber \\
        &= 4\pi\,\rho_{\mathrm{s}}\,r_{\mathrm{s}}^3 
           \left[\ln(1+c_{200}) - \frac{c_{200}}{1+c_{200}}\right].
\end{align}
Substituting $\rho_{\mathrm{s}} = \kappa_{\mathrm{s}}\,\Sigma_{\mathrm{crit}}/r_{\mathrm{s}}$ yields the implicit relation
\begin{equation}
\frac{\kappa_{\mathrm{s}}\,\Sigma_{\mathrm{crit}}}{r_{\mathrm{s}}}
\left[\ln(1+c_{200}) - \frac{c_{200}}{1+c_{200}}\right]
= \frac{200}{3}\,\rho_{\mathrm{crit}}(z_{\mathrm{l}})\,c_{200}^3,
\end{equation}
which can be solved numerically for $c_{200}$, with $r_{200}$ and $M_{200}$ following directly. The halo parameters $(M_{\rm vir}, r_{\rm vir})$ can be obtained analogously, using the virial overdensity $\Delta_{\rm vir}(z_{\rm l})$ at the lens redshift. The mass enclosed within $r_{\rm vir} = c_{\rm vir}\, r_{\mathrm{s}}$ is
\begin{equation}
\begin{split}
M_{\rm vir} &= \frac{4\pi}{3}\,\Delta_{\rm vir}(z_{\rm l})\,\rho_{\rm crit}(z_{\rm l})\,r_{\rm vir}^3 \\
            &= 4\pi\,\rho_{\mathrm{s}}\,r_{\mathrm{s}}^3 
               \left[\ln(1+c_{\rm vir}) - \frac{c_{\rm vir}}{1+c_{\rm vir}}\right].
\end{split}
\end{equation}

We estimate that the detected subhalo has a mass of $\rm M_{200} = 2.78_{-0.66}^{+0.43} 10^{10} \, \Msol$ and a concentration of $c_{200} = 30_{-7}^{+5}$ (see top panel in Figure~\ref{fig:enclosed_mass}). This concentration is consistent with that expected for a typical NFW subhalo of this mass that has been tidally stripped inside the parent halo \citep{Springel_2008}\footnote{According to \citealt{Springel_2008} tidal stripping typically increases the characteristic density of an NFW subhalo by a factor of 2.6, corresponding to an increase in concentration by a factor of 1.5 relative to a field halo of the same mass, in this case, $c\simeq 12.7$.}.
At the virial radius, the parameters become $\rm M_{\rm vir} = {2.93}_{-0.71}^{+0.46} \times 10^{10} \, \Msol$ and $c_{\rm vir} = 34_{-8}^{+5}$. 

For comparison, the two confirmed subhalo detections in the literature, SDSS~J0946+1006 and JVAS~B1938+666, have $\rm M_{\rm vir} = {1.78}_{-0.32}^{+0.36} \, 10^{10} \, \Msol$ and $c_{\rm vir} = 250_{-53}^{+78}$, and $\rm M_{\rm vir} = {5.08}_{-2.20}^{+3.96} \, 10^{8} \, \Msol$ with $c_{\rm vir} = 185_{-65}^{+116}$, respectively \citep[][]{2025MNRAS.543..540T}. We note that all existing detections that treat concentration as a free parameter lie above the median mass--concentration relation (MCR). As discussed in \citet{2022MNRAS.510.2464A}, more concentrated subhalos produce stronger lensing perturbations and are therefore more readily detectable, implying a natural observational bias towards higher inferred concentrations in detected systems. 

As seen in the left and middle panels of Figure~\ref{fig:temp_1}, the parameters $\kappa_{\rm s}$ and $r_{\rm s}$ are strongly degenerate, such that different combinations of these quantities produce nearly identical lensing effects. Consequently, the inferred value of $M_{200}$ (or $M_{\rm vir}$) is not well constrained and cannot be regarded as a reliable estimate of the subhalo’s total mass. Following the definition introduced by \cite{2025MNRAS.543..540T}, we instead identify the robust radius, $R_{\rm sub}$, as the radius at which the relative uncertainty on the deflection angle (equivalently on the projected enclosed mass) is minimized for the best-fitting model \citep[see also][]{2021MNRAS.507.1662M}. This radius marks the spatial scale where the lensing signal of the subhalo is most tightly constrained by the data and therefore provides a more meaningful measure of its mass than the extrapolated $M_{200}$ value.

In Figure~\ref{fig:enclosed_mass} we show the projected enclosed mass profiles, $M( < R)$, for the two subhalo models that were used to fit the data. Since both models result in the same log Bayesian evidence, we choose the model with fewer parameters, i.e. the NFW profile, to compute the robust radius (although both give similar values). The fractional error on the enclosed mass is minimum at a radius of, $R_{\rm sub} \sim 2$~kpc, and the mass enclosed within that radius is $M_{\rm sub} = {3.57}_{-0.14}^{+0.16} \, 10^9 \Msol$. For comparison, the robust radii for the two previous detections are $R_{\rm p} \approx 700$~pc for SDSS~J0946+1006 and $R_{\rm p} \approx 90$~pc for JVAS~B1938+666 \citep[][]{2025MNRAS.543..540T}.

%%%%%%%%%%%%%%%%%%%%%%%%%%%%%%%%%%%%%%%%%%%%%%%%%%%%%%%%%%%%%%%%%%%%%%%%%%%
% SECTION
%%%%%%%%%%%%%%%%%%%%%%%%%%%%%%%%%%%%%%%%%%%%%%%%%%%%%%%%%%%%%%%%%%%%%%%%%%%
\section{Conclusions} \label{sec:section_6}

In this work, we have presented a detailed strong lensing analysis of the system PJ011646, using ALMA observations of the dust continuum emission. Our aim is to test whether the data favour the presence of a dark matter subhalo in this lensing system. Our main conclusions are summarised as follows.

We find that the data strongly support the inclusion of multipole terms of orders $m = 1, 3,$ and $4$ in the total mass-density distribution of the lens. The corresponding amplitudes are $\sim$1.5~percent of the convergence for the third- and fourth-order multipoles, and $\sim$3.5~percent for the first-order term. These values are consistent with expectations from studies of local early-type galaxies, which show similar multipole amplitudes in their stellar light distribution based on isophotal analyses \citep{2018ApJ...856...11G, 2025MNRAS.540.3281A}. The inclusion of these components provides the flexibility required to capture low-level angular structure in the mass distribution and prevents spurious detections of small-scale perturbations that might otherwise mimic the signal of a dark matter subhalo \citep[e.g.][]{2024MNRAS.528.1757O}.

After accounting for the angular structure in the macromodel, we find evidence for a dark matter subhalo of mass $\rm M_{200} = 2.78_{-0.66}^{+0.43} 10^{10} \, \Msol$ and  concentration $c_{200} = 30_{-7}^{+5}$, assuming it lies at the redshift of the lens. This concentration is consistent with that expected for an NFW subhalo after it has been stripped in its host halo. The subhalo is located $\delta x = 2.07 \pm 0.03$ and $\delta y = -0.88 \pm 0.03$~arcsec away from the centre of the main lens galaxy. The model including this subhalo is preferred over the pure PL+MP ($m = 3,4$) macromodel with a log-evidence difference of $\Delta \ln Z = 22.5$, corresponding to a detection significance of $\sim5.8\sigma$. The evidence increase of the PL+MP model with multipole terms $m = 1, 3, 4$ relative to the $m = 3, 4$ model is $\Delta \ln Z = 6.5$, which is significantly smaller than the improvement obtained by including a subhalo. This indicates that the addition of the $m = 1$ multipole alone cannot account for the evidence gain associated with the subhalo. For this reason, we do not repeat the subhalo search using the $m = 1, 3, 4$ macromodel.

We estimate the radius of the detected subhalo to be $\theta_{\rm sub} = 0.29$~arcsec (equivalent to $R_{\rm sub} \sim 2$~kpc at the redshift of the lens), defined as the radius where the fractional uncertainty on the projected enclosed mass is minimised. The corresponding enclosed projected mass within this radius is $M_{\rm sub, \, 2D} = {3.57}_{-0.14}^{+0.16}\times 10^9 \Msol$. This scale corresponds to the radius at which the lensing signal of the subhalo is most tightly constrained by the data and provides the most reliable physical quantity for comparison with other detections and theoretical predictions.

We additionally tested a generalised NFW (gNFW) profile for the subhalo in the same grid cell where the detection was identified. The resulting model yields an equivalent log Bayesian evidence to that of the NFW profile, suggesting that the data do not require a deviation from the standard NFW form within the range of scales probed. We were able to place an upper limit, $\gamma_{\rm gNFW} < 1.7$, at $3\sigma$ on the slope of the gNFW profile. 

Finally, using grid cells consistent with no subhalo detection ($\Delta \ln Z \simeq 0$), we derive upper limits on subhalo properties across the image plane. The minimum detectable subhalo mass at a $3\sigma$ significance level is $M_{200, \rm min} \approx 8 \times 10^{8} \, \Msol$, assuming the subhalo lies on the MCR from \citet{2016MNRAS.460.1214L}. These limits represent the best achievable constraints for subhalos located along the bright lensed arcs, where the data are most sensitive; away from the arcs, the constraints become progressively weaker. Although the angular resolution of our data ($\rm FWHM \approx 0.1$~arcsec) is comparable to that of HST and not the finest achievable with ALMA, the high magnification of this system enables us to probe subhalo masses below $10^9 \, \Msol$.

Building on the results presented here, we will extend this analysis to the growing sample of strongly lensed DSFGs observed with ALMA at sub-50~mas resolution. This will enable systematic, statistically robust searches for dark matter substructure across the population, pushing the sensitivity of lensing-based subhalo detections to masses $M_{200}\lesssim 10^{8} \, \Msol$ \citep[e.g.][]{2025arXiv250802776S}. Such measurements will provide a strong test of the small-scale predictions of the $\Lambda$CDM model and open a new observational window on the nature of dark matter.

%%%%%%%%%%%%%%%%%%%%%%%%%%%%%%%%%%%%%%%%%%%%%%%%%%%%%%%%%%%%%%%%%%%%%%%%%%%
% SECTION
%%%%%%%%%%%%%%%%%%%%%%%%%%%%%%%%%%%%%%%%%%%%%%%%%%%%%%%%%%%%%%%%%%%%%%%%%%%
\section*{Acknowledgements}

AA, QH, SMC and CSF acknowledge support from the
European Research Council (ERC) Advanced Investigator grant DMIDAS (GA 786910), to C.S. Frenk. XYC acknowledges the support of the National Natural Science Foundation of China (No.\ 12303006). This paper makes use of the DiRAC Data-Centric system, project codes dp004, which is operated by the Institute for Computational Cosmology at Durham University on behalf of the STFC DiRAC HPC Facility (www.dirac.ac.uk). The DiRAC facility was funded by BIS National E-infrastructure capital grant ST/K00042X/1, STFC capital grants ST/H008519/1, ST/K00087X/1, ST/P002307/1, ST/R002425/1, the STFC DiRAC Operations grant ST/K003267/1, and Durham University. DiRAC is part of the UK National E-Infrastructure.

This paper makes use of the following ALMA data: ADS/JAO.ALMA \# 2022.1.01311.S ALMA is a partnership of ESO (representing its member states), NSF (USA) and NINS (Japan), together with NRC (Canada), MOST and ASIAA (Taiwan), and KASI (Republic of Korea), in cooperation with the Republic of Chile. The Joint ALMA Observatory is operated by ESO, AUI/NRAO and NAOJ.

\section*{DATA AVAILABILITY}
All data used in this work are publicly available.

\section*{Software Citations}

This work uses the following software packages: \href{https://github.com/numpy/numpy}{{NumPy}}, \href{https://github.com/matplotlib/matplotlib}{{Matplotlib}}, \href{https://github.com/astropy/astropy}{{Astropy}}, \href{https://github.com/scipy/scipy}{{Scipy}}, \href{https://github.com/jyhmiinlin/pynufft}{PyNUFFT}, \href{https://github.com/jyhmiinlin/pynufft}{Nautilus}

%%%%%%%%%%%%%%%%%%%%%%%%%%%%%%%%%%%%%%%%%%%%%%%%%%%%%%%%%%%%%%%%%%%%%%%%%%%%%
%%%%%%%%%%%%%%%%%%%%%%%%%%%%%%%%%%%%%%%%%%%%%%%%%%%%%%%%%%%%%%%%%%%%%%%%%%%%%

\bibliographystyle{mnras}
\bibliography{main}

%%%%%%%%%%%%%%%%%%%%%%%%%%%%%%%%%%%%%%%%%%%%%%%%%%%%%%%%%%%%%%%%%%%%%%%%%%%%%
%%%%%%%%%%%%%%%%%%%%%%%%%%%%%%%%%%%%%%%%%%%%%%%%%%%%%%%%%%%%%%%%%%%%%%%%%%%%%
% \appendix

%%%%%%%%%%%%%%%%%%%%%%%%%%%%%%%%%%%%%%%%%%%%%%%%%%%%%%%%%%%%%%%%%%%%%%%%%%%%%%
%%%%%%%%%%%%%%%%%%%%%%%%%%%%%%%%%%%%%%%%%%%%%%%%%%%%%%%%%%%%%%%%%%%%%%%%%%%%%%

\end{document}